\documentstyle[prl,aps,psfig,twocolumn]{revtex}  % REVTeX v.3.0. For galley proofs

\newcommand \be {\begin{equation}}
\newcommand \bea {\begin{eqnarray} \nonumber }
\newcommand \ee {\end{equation}}
\newcommand \eea {\end{eqnarray}}

\newcommand{\bit}{\begin{itemize}}
\newcommand{\eit}{\end{itemize}}

\begin{document}
\setcounter{page}{1}
\draft

\title{On the Shape of the Tail of a Two Dimensional Sand Pile} 
\author{ J.J.Alonso and H.J.Herrmann}
\address{
Laboratoire P.M.M.H.
\\ \'Ecole Sup\'erieure de Physique et Chimie Industrielles \\ 
10 rue Vauquelin 75005 Paris}

\maketitle
\begin{abstract}

We study the shape of the tail of a heap of 
granular material. A simple theoretical
argument shows that the tail adds a logarithmic
correction to the slope given by the angle of repose. This
expression is in good agreement with experiments. We present a 
cellular automaton that contains gravity, dissipation 
and surface roughness and its simulation also
gives the predicted shape.

\end{abstract}

\pacs{PACS numbers: 64.75.+g, 05.70.Jk, 46.10.+z}

\narrowtext   
%%%%%%%%%%%%%%%%%%%%%%%%%%%%%%%%%%%%%%%%%%%%%%%%%%%%%%%%%%%%%%%%%%%
In this letter we discuss the shape of static heaps of granular
material. In fact many typical phenomena observed in granular
materials have been studied in the last years~\cite{WIL,HW,RSPS}. 
In order to understand these phenomena simple models as well as
computer intensive simulation techniques~\cite{HJH} 
have been very useful.

As everybody knows real sand piles are almost perfect cones
with  a well defined angle
of repose which depends on gravity and on the characteristics of the 
material. Watching carefully, however, one notices
the existence of curved tails at the bottom of the heaps
to which in fact not much attention has been paid in the literature
up to now. In the following we shall study the shape of 
these tails in more detail.

We will present 
a simple experiment to obtain the profile of the tails in two
dimensions. Then, we deduce an analytical 
expression for their shape
which agrees with our experimental profiles.
This expression has two parameters that describe the
material properties.
Finally, we test the theory with data 
obtained from computer simulations using cellular automata (CA).

Several CA models have been
proposed in the past to study the
behaviour of granular matter \cite{Baxter,KK,DMar,PH} .
They include dissipation which is  
the most important ingredient to capture the peculiarities
of granular media. One such automaton 
is a lattice gas formulation by Peng et al.~\cite{PH} with
rest particles and inelastic collision rules. 
Using this model we have simulated 
two dimensional heaps in a 
quasi-static regime and compared the resulting profiles with 
those predicted by our theory.

We built sand piles in an easy and inexpensive experiment.
Grains were poured at a slow rate of about ten particles per second 
from the top into the center of a
rectangular cell made of two  parallel vertical glass plates
of size $30 \times 25~ cm^{2}$ separated by a fixed distance of 2~$mm$. 
As granular materials we used lead spheres, sugar and polenta.
The grain diameter was 2~$mm$ in the first case and about 0.5~$mm$
in the other two cases. In fig.~1(a) we see a digizited image
of the experiment showing the essentially two-dimensional
heaps that we obtain at different times. 
We have studied the profiles of the heaps
by recording digitized pictures taken by a VHS video camera at
different stages of the evolution. The resolution of digitized 
images was 40 pixels/$cm$.

The heaps are grown in steady state, in the sense that
only a few grains move simultaneously being in the constant
velocity regime~\cite{Riguidel}.
In this regime particles do not accumulate at the top of the heap 
and the formation of big avalanches which can
disturb the formation of smooth profiles is not present. 
Therefore profiles do not depend on the flow rate 
at the top. They only depend on the type of grains used
and gravity. 

In Fig.~1(a) we see the typical situation of two flat
surfaces that define the angle of repose. 
We will, however, focus our attention on the small curved 
tails that can be observed at the base of the heaps and 
obtain the analytical expression of their shape using a 
simple argument.

Taking pictures at different times we discovered that
it is possible to superpose the profiles of the
tails by a simple horizontal shift (see Fig.~1(a)).
It seems that the heap grows by
putting layers of particles over one another. 
Each individual layer grows upwards 
from the bottom to the top by stopping
particles which are moving down along the surface of the heap. 

\renewcommand{\thefootnote}{\fnsymbol{footnote}}

At the top of each incomplete layer there is a kink
(kinks are marked by arrows in Fig.~1(b)){\footnote {In 
the case of irregulary shaped particles or particles of 
different size 
one has asperities of different size on the surface of the
heap and the larger ones effectively act as the kinks discussed here. }
The presence of such
kinks reduces the slope of the surface away from 
the angle of repose of the material. 
Let us describe the surface by $h(x)$
where $h$ is the height and $x$ the corresponding horizontal 
displacement. We choose the origin at the center of the base of the heap, i.e.
$h(0)=h_m$ where $h_m$ is the height of the heap at the top, and consider only
$x~>~0$ since the heap is symmetric with respect
to the origin. In the absence of kinks 
one would have $h = h_m - \gamma x$ where 
$\gamma = tan ~\theta$ and $\theta$ is the angle of repose. 
The presence of each kink increases this ideal value of $x$ by
a certain value $l$ which typically is of the size of a grain. If $\rho$ is the
number of kinks per unit length in the vertical direction
we can express the slope of the surface as a function of $\rho$
as:
\be
\frac {dh}{dx} = \frac {- \gamma } {1+l \gamma \rho }
\ee
Particles falling down along
the surface collide with the kinks and can be
accumulated on top of them with some rate $r$
or continue to move downwards to the next layer.
$r$ depends on $l$
and on the properties of the 
grains (shape, roughness, coefficient of restitution).
Let us call $\Phi(h)$ the flux of particles pulled
down by gravity along the surface. Since the experiments 
were performed in steady state one has the relation
$\frac {d \Phi}{dh} = r \Phi \rho$.
The fact that the heap grows by a shift of the profile
in the horizontal direction means that the aggregation rate of 
particles is independent of $h$. 
Since the number of particles aggregated per vertical unit length is
the variation of the vertical flux we have  $\partial \Phi/\partial h = B$,
where $B$ is a constant and since $\Phi(0)=0$ we obtain $\Phi = Bh$.
Putting everything together the slope of the surface
as a function of $\rho$ is given by:
\be
\frac {dh}{dx} = \frac {-  \gamma}{1+\frac{l \gamma}{r h}}
\ee
After integration we obtain the final expression,
\be
x = \frac{h_m - h}{\gamma} + l_e~ log \frac{h_m}{h}
\ee
which shows a logarithmic correction to the angle of repose. 
$\gamma = tan ~\theta$ ~is the classical dynamic friction coefficient
and $l_e\equiv \frac l{r}$ ~describes the typical extra horizontal
displacement a particle must undergo before it sticks.
Both constants characterize the granular material.
In Figure~1(c) we show the average over 10 different profiles representing
the deviation from the straight profile. Error bars are the standard 
deviations for each value of $h$. The surface fit is for $\gamma=0.98$
and $l_e= 1.5 mm$. It is very interesting to note that for the
three materials used here we find consistence with  $r \approx 1/3$.

In the following we will check our formula using numerical
simulations. We consider a Lattice Gas Automaton (LGA) at integer 
time steps $t=0,1,2,...$ with particles
located at the sites of a two dimensional triangular lattice of
size $L$. Gravity acts downward in the vertical direction and 
forms an angle of $30^{0}$ with the closest lattice axis.
At each
site there are seven bit variables which refer to the velocities
$v_i (i=0,1,2,...6)$. Here $v_i (i=1,...6)$ are the nearest neighboring (NN)
lattice vectors and $v_0=0$ refers to the rest state (zero velocity). Each
state can be either occupied by a single particle or empty. Therefore, the
number of particles per site has a maximal value of seven and a minimal value
of zero. 

The time evolution of the LGA consist of a collision step and a 
propagation step. In the collision step particles can change their 
velocities due to collisions and in the propagation step particles move
in the direction of their velocities to the NN sites where they collide again.
The system is updated in parallel. Only the collisions specified
in Figure~2 can deviate the trajectories of particles 
from straight lines with probabilities depending on the 
dissipation parameter $p$. 
If $p$ is not zero, energy can be dissipated 
during the collision. This is a crucial property of granular
materials and yields among other an instability towards
cluster formation~\cite{GOZA}.

The two collisions shown 
on the lower part of Fig.~2 temporarily allow 
more than one particle on a site. However, immediately after 
the collision step, the extra rest particles
randomly hop to NN sites until they find a site with no rest
particle and there they stop.

We incorporate the driving force, namely gravity $g$, by
setting a rest particle into motion with probability $g/2$ along 
one of the two lattice directions which form an angle
of $30^{0}$ with the direction of gravity, however, only
if the site below is empty at that time.  Rest particles that are
already on the heap can only be accelerated by gravity if at least
one of the two NN sites in the direction of gravity does not yet
belong to the heap.

During the evolution, we add particles at
a fixed rate from one site at the top to the system. Gravity moves
these particles down until they collide with the
hard wall at the bottom. A particle colliding with the bottom is either
reflected with probability $1-p$ in the specular direction or 
stopped with probability $p$ losing its momentum.  
In the second case the resulting rest particle belongs to
the growing heap, and we label that particle in order to store the
information that it is sustained by the bottom plate.
Every particle colliding with
these particles looses its momentum and is aggregated to the
heap after the redistribution process described above
has taken place.
 
We perform simulations for systems of size $L=512$ 
typically iterating $3 \times 10^5$ time steps. Most of the computer work was 
performed on a CM-5. Like in the experiment we add few particles
(one every 8 time steps) to be in a quasi-static regime.
In that case the profile does not depend 
on the history of the system. 

For all parameter values the profiles have an
angle of repose of $60^{0}$ corresponding to $\gamma = \sqrt{3}$ ~which
is determined by the geometry of the underlying lattice. The tail
shows the presence of kinks which reduce the slope. The different 
gray levels correspond to different time steps. One notes that 
like in the experiment the surfaces of the heap are just horizontally 
displaced in time.

For this model the parameters in equation~(3) are easily determined. 
On one hand we use as unit length the distance between NN sites on 
the underlying lattice which we choose to be unity
giving~$l=1$. On the other hand $r=1/3$
because every particle colliding with a kink is aggregated to the 
heap and then redistributed randomly to one of the three 
empty NN sites. Only one of these sites particles will aggregate
on top of a layer.

In Figure~3 we have plotted an average of 16 independent
profiles in the same manner as in Fig.~1(b). The continous line 
is the shape resulting from  equation~(3) with $\gamma = \sqrt{3}$ 
and $l_e = 3$. Our formula is in excellent quantitative agreement 
with the simulation. 

In order to include into the CA the sticking of particles typical
for wet sand we 
introduce a new parameter, namely 
the probability $\eta$ that one particle stopped and aggregated 
after a collision with the heap is
``blocked''. Gravity can only move a ``blocked'' particle if none 
of the two NN sites below belongs to the heap and then
the particle is no longer 
blocked. Introducing these rules the slope of the surfaces of the 
heaps can be larger 
than $60^{0}$ because each blocked particle can support rest particles.

Averaged profiles for $\eta=0.002$ and $0.004$ are shown in Fig.~4. 
We observe that now there is no well defined angle of repose like 
in the case of dry sand. The profile can be calculated using very 
similar arguments to 
those used before.

Let $\rho_0$ be  the vertical density of blocked particles on the surface.
The relation between the slope of the profile and $\rho_0$ is then 
\be
\frac {dh}{dx} = \frac { - \sqrt{3} } {1+ \sqrt{3}~ (\rho - \rho_0) }
\ee
because the presence of each blocked particle on the surface reduces the 
value of $x$ by $l=1$.

It is easy to find a relation between $\rho$ and $h$.
On one hand  $\rho_0 \propto \eta \Phi$.
On the other hand $\Phi \propto h$ due to the
translational invariance which we have observed experimentally 
so that $\rho_0 = c' \eta h$.
One therefore obtains,
\be
\frac {dh}{dx} = \frac {- \sqrt{3}} {1+ \sqrt{3} ( \frac 1 {r h} - c' \eta h) }
\ee
which after integration gives 
\be
x = \frac{h_m - h}{\sqrt{3}} (1 - c(h+h_m)) + r^{-1}~ log \frac{h_m}{h}
\ee
where $c=\frac {\sqrt{3}}{2} c' \eta$ 
is a constant.  In Fig.~4 we also show the fits obtained from 
this expression using $c'=0.39$ and $r=0.35$.
We have checked that $\rho_0 \propto h$
in agreement with the argument for $h<100$. If $h$ becomes larger
than 100 the density of blocked particles saturates to $\rho_0 \approx
0.28$.

We have studied experimentally, theoretically and numerically with a
cellular automaton the tail of sand piles. The theoretical argument
giving the shape of the tail is based on the experimentally observed
translational invariance and uses mass conservation. One finds that the tail
constitutes a logarithmic correction to the naive straight slope given
by the angle of repose. The analytic expression fits very well to the
experimental shape and to the two-dimensional heaps obtained
with a cellular automaton. 

\vskip 1cm
\subsubsection*{Acknowledgments}

We are indebted to S. Roux for fruitful discussions,
to P.Petitjeans and D.Hoang for help on the experimental
work and to H. Puhl for much help on the computer.
JJA thanks the CNCPST for a generous grant of computer time on the CM-5
and is grateful for an {\it Ayuda parcial} (PB91-0709)
and a postdoctoral grant from DGICYT.

\newpage
\subsubsection*{Figure captions}
\begin{list} {}{\itemindent -2.5cm \leftmargin 2.7cm \labelsep 0cm
                \listparindent 0cm \labelwidth 2.5cm}

\vspace{0.4cm}
 
\item{\makebox[2.5cm][l]{[Fig.~1a, b, c]}}
(a) Digitized image of a heap of the polenta. The diameter of the
grains is about 0.5 $mm$. The height of the heap is 16 $cm$.
Different gray levels show the pile at different stages of
growth. The superposed continous lines in both figures are fits 
obtained from eq.\ (3) by taking the values $\gamma=0.98$ 
and $l_e=1.5~mm$.
(b) Tail of a pile made of lead spheres with a diameter of 2 $mm$.
One can observe parallel layers terminating in
horizontal kinks (marked by arrows).
(c) Deviation of the shape from the straight profile 
for the same material. Data are averages obtained from 10 
different samples. Error bars represent the standard 
deviations of the averaged values. Continuous line represent 
the same fit of Fig.~(a) 

\item{\makebox[2.5cm][l]{[Fig.~2]}}
Collision rules of the cellular automaton. Arrows represent moving particles
and full dots stand for rest particles. The number next to each 
configuration is the probability for that transition.

\item{\makebox[2.5cm][l]{[Fig.~3]}}
Deviations from from the straight line for a heap of 80000 
particles obtained in a quasi-static regime adding
one particle every 8 time steps. 
Data are averages obtained from
16 independent samples. The fit is the same as in fig~3(a)

\item{\makebox[2.5cm][l]{[Fig.~4]}}
Shapes obtained from 16 independent samples for a roughness parameter 
$\eta=0.002$ (circles) and $\eta=0.005$ (squares). Errors are of the
same size of symbols. Continuous lines are fits obtained from 
eq.\ (6) with $c'=0.39$ and $r=0.35$. 

\end{list} 

\begin{thebibliography}{99}


\bibitem{WIL}  H.M. Jaeger and S.R. Nagel, Science {\bf 255},
1523 (1992).

\bibitem{HW} A. Hansen and D. Bideau (eds.), {\it Disorder and
Granular Media} (North Holland, Amsterdam, 1992).

\bibitem{RSPS} A. Mehta (ed.), {\it Granular Matter} (Springer,
Heidelberg, 1994).

\bibitem{HJH} H.J. Herrmann, III Granada School, Lecture Notes in Physics, 
vol. 448, Springer Verlag (1994).

\bibitem{Baxter} G.W. Baxter and R.P. Behringer, Phys. Rev A
{\bf 42}, 1017 (1990); Physica D {\bf 51}, 465 (1991).

\bibitem{KK} A. K\'arolyi, and J. Kert\'esz, Proceedings of the 6th
Joint EPS-APS International Conference on Physics Computing, (1994);
S. Vollmar and H.J.Herrmann, preprint.

\bibitem{DMar} D. D\'esirable and J. Martinez, in {\it Powder and Grains},
ed. C. Thornton (Balkema, Rotterdam, 1993), p.345;

\bibitem{PH}  G. Peng and H.J.Herrmann, Phys.Rev. E {\bf 48},
R1796 (1994) and preprint.
  
\bibitem{Riguidel} G.H. Ristow, F.-X. Riguidel and D. Bideau,
J. Physique I, {\bf 4}, 1161 (1994).

\bibitem{GOZA} I. Goldhirsch and Z. Zanetti, Phys. Rev. Lett.
{\bf 70}, 1619 (1993).

 \end{thebibliography}
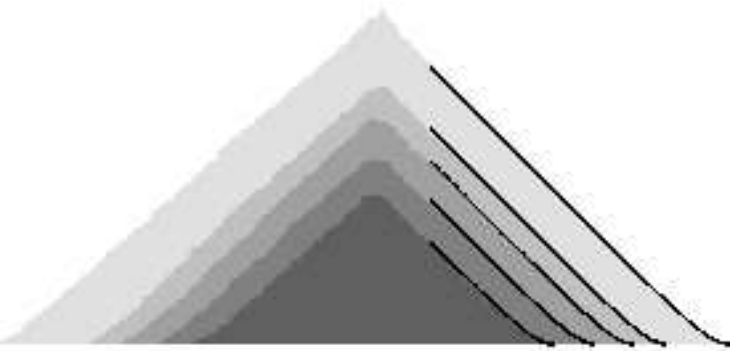
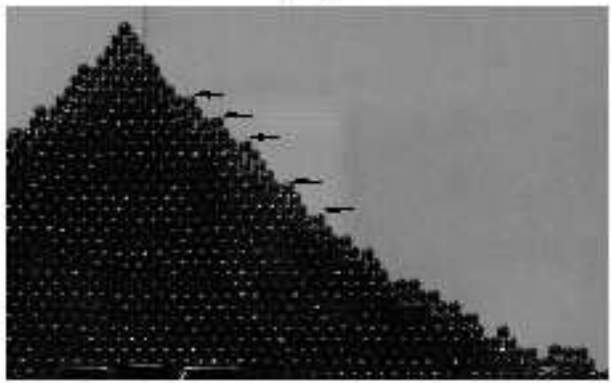
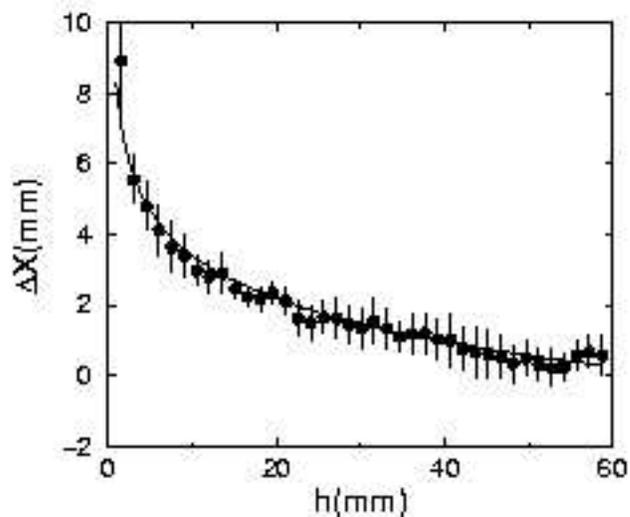
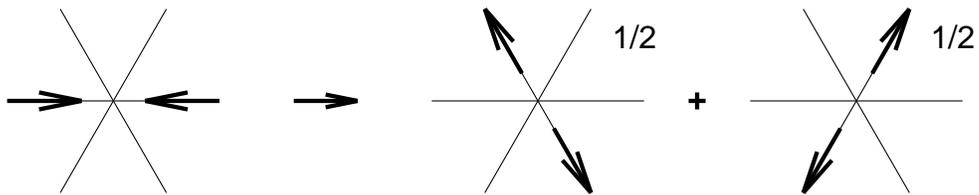
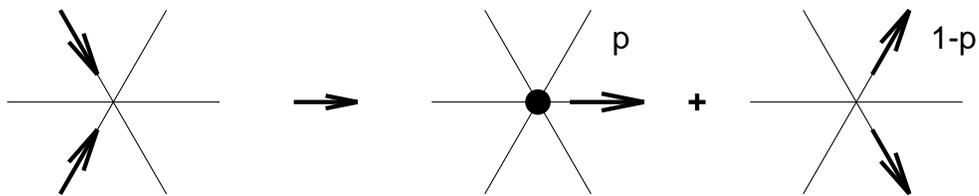
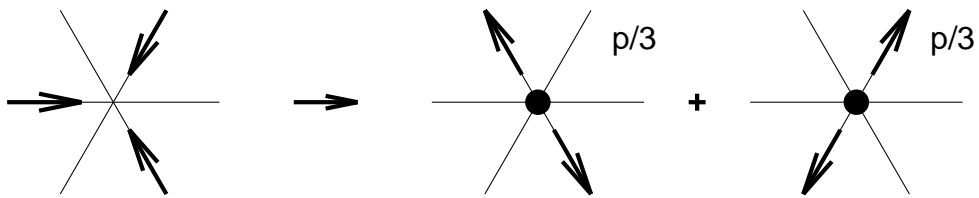
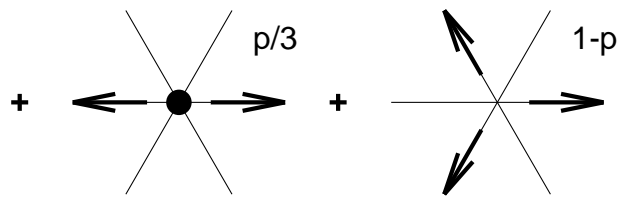
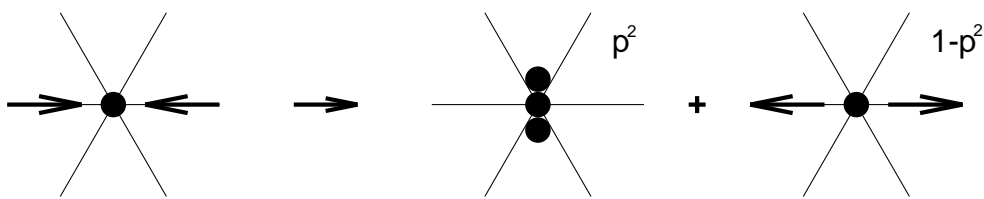
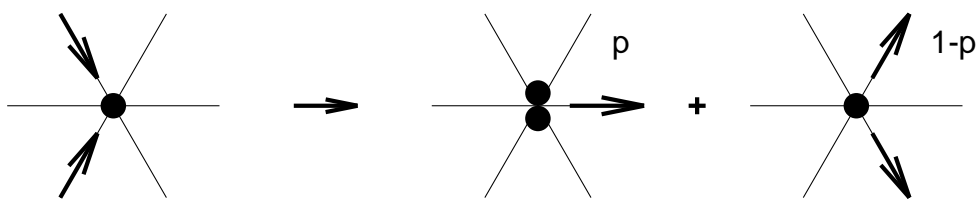
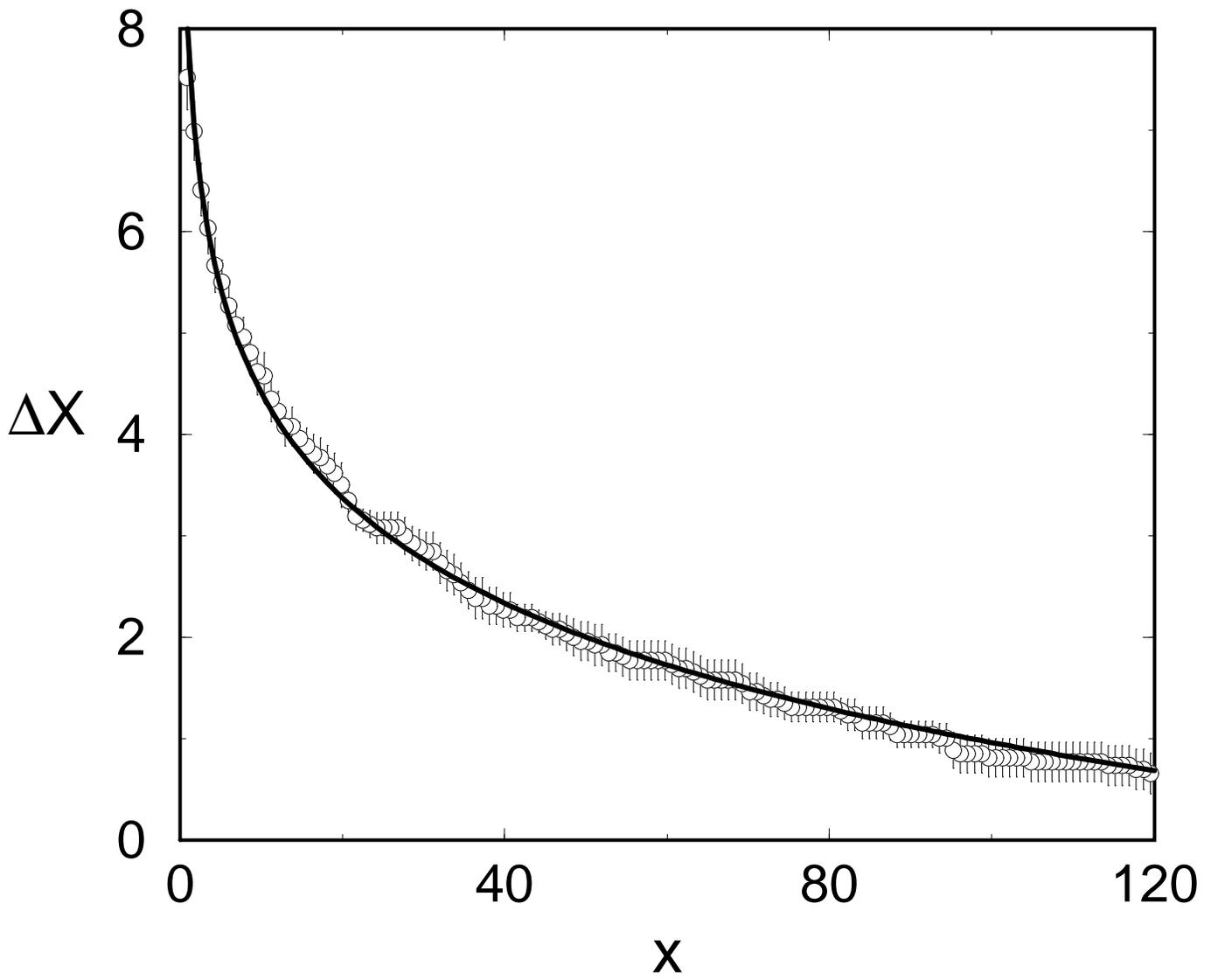
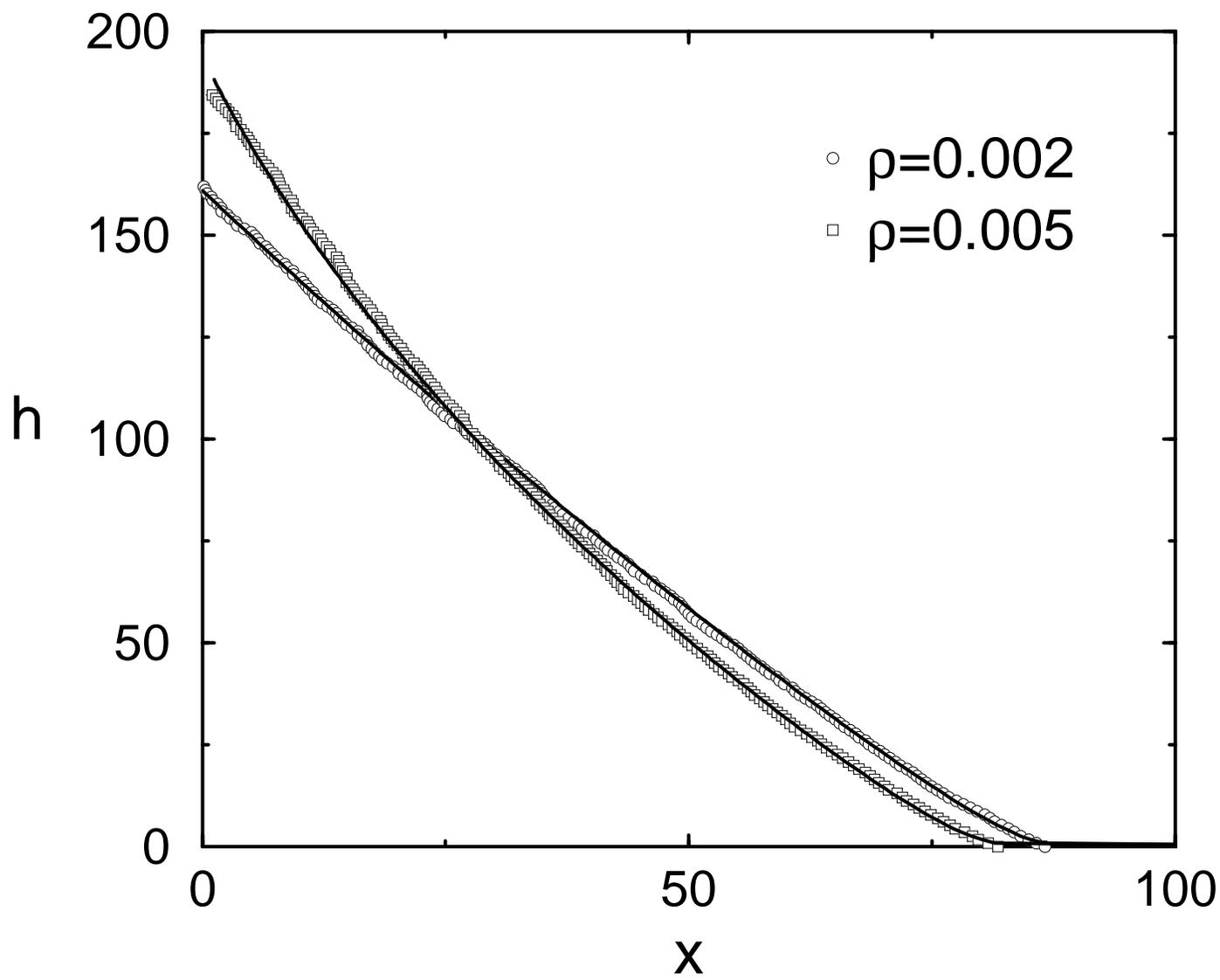
\end{document}